\documentclass{emulateapj}
\usepackage{graphicx}
\usepackage{times}

\newcommand{\feh}{{\rm [Fe/H]}}
\def\gsim { \lower .75ex \hbox{$\sim$} \llap{\raise .27ex \hbox{$>$}} }
\def\lsim { \lower .75ex \hbox{$\sim$} \llap{\raise .27ex \hbox{$<$}} }
\shorttitle{Where are the very metal-poor stars?} 
\shortauthors{A. Helmi et al.}

\begin{document}

\title{A new view of the dwarf spheroidal satellites of the Milky Way
from VLT/FLAMES\altaffilmark{1}: Where are the very metal poor stars?}

\author{Amina Helmi\altaffilmark{2}, M.J. Irwin\altaffilmark{3},
E. Tolstoy\altaffilmark{2}, G. Battaglia\altaffilmark{2},
V. Hill\altaffilmark{4}, P. Jablonka\altaffilmark{5},
 K. Venn\altaffilmark{6},
M. Shetrone\altaffilmark{7}, B. Letarte\altaffilmark{2},
N. Arimoto\altaffilmark{8}, T. Abel\altaffilmark{9},
P. Francois\altaffilmark{4,}\altaffilmark{10}, %%% note change {3,10}
A. Kaufer\altaffilmark{10}, F. Primas\altaffilmark{11},
K. Sadakane\altaffilmark{12}, T. Szeifert\altaffilmark{10} }

\altaffiltext{1}{Based on FLAMES/GIRAFFE observations
collected at the European Southern Observatory,
proposal 171.B-0588}
\altaffiltext{2}{Kapteyn Astronomical Institute, University of Groningen, 
P.O.Box 800, 9700 AV Groningen, The Netherlands;  
{\sf{ahelmi@astro.rug.nl}}}
\altaffiltext{3}{Institute of Astronomy, University of Cambridge,
Madingley Road, Cambridge CB3 0HA, UK}
%; {\sf{mike@ast.cam.ac.uk}}}
\altaffiltext{4}{Observatoire de Paris, section de Meudon,
5 Place Jules Janssen, F-92195 Meudon Cedex, France}
%; {\sf{Vanessa.Hill, patrick.Francois@obspm.fr}}}
\altaffiltext{5}{Laboratoire d'Astrophysique, Ecole Polytechnique Federale de
Lausanne (EPFL), Observatoire, CH-1290 Sauverny, Switzerland.
%; {\sf{Pascale.Jablonka@obs.unige.ch}}}
%\altaffiltext{6}{
On leave from CNRS-UMR8111, Observatoire de Paris}
\altaffiltext{6}{Department of Physics and Astronomy,
University of Victoria, 3800 Finnerty Road, Victoria, BC, V8P 1A1, Canada}
%; {\sf{kvenn@uvic.ca}}}
\altaffiltext{7}{University of Texas, McDonald Observatory, Fort
Davis, Texas, USA}
%; {\sf{shetrone@shamhat.as.utexas.edu}}}
\altaffiltext{8}{National Astronomical Observatory of Japan, 2-21-1 Osawa, 
Mitaka, Tokyo 181-8588, Japan}%; {\sf{arimoto@optik.mtk.nao.ac.jp}}}
\altaffiltext{9}{Kavli Institute for Particle Astrophysics and Cosmology,
%Department of Physics and Stanford Linear Accelerator Center, Stanford
%University, 
2575 Sand Hill Road, Menlo Park, CA 94044, USA}%; 
%{\sf{tabel@slac.stanford.edu}}}
\altaffiltext{10}{European Southern Observatory, Alonso de Cordova 3107, 
Santiago, Chile}
%; {\sf{akaufer, tszeifer@eso.org}}}
\altaffiltext{11}{European Southern Observatory,
Karl-Schwarzschildstr. 2, D-85748 Garching bei M\"{u}nchen, Germany}
%; {\sf{fprimas@eso.org}}}
\altaffiltext{12}{Astronomical Institute, Osaka Kyoiku University,
Asahigaoka, Kashiwara, Osaka 582-8582, Japan}
%; {\sf{sadakane@cc.osaka-kyoiku.ac.jp}}}

\begin{abstract}
As part of the Dwarf galaxies Abundances and Radial-velocities Team
(DART) Programme, we have measured the metallicities of a large sample
of stars in four nearby dwarf spheroidal galaxies (dSph): Sculptor,
Sextans, Fornax and Carina.  The low mean metal abundances and the
presence of very old stellar populations in these galaxies have
supported the view that they are fossils from the early Universe.
However, contrary to naive expectations, we find a significant lack of
stars with metallicities below $\feh \sim -3$ dex in all four
systems. This suggests that the gas that made up the stars in these
systems had been uniformly enriched prior to their formation.
Furthermore, the metal-poor tail of the dSph metallicity distribution
is significantly different from that of the Galactic halo. These
findings show that the progenitors of nearby dSph appear to have been
fundamentally different from the building blocks of the Milky Way,
even at the earliest epochs.
\end{abstract}


\section{Introduction}
\label{sec:intro}

The relevance of dwarf galaxies in the Local Group is two-fold. As
nearest neighbour systems, they can be studied in great detail and
hence they are excellent cases for understanding star formation and
prototypical galaxy evolution over the lifetime of the Universe (Mateo
1998). Furthermore, dwarf galaxies are the simplest systems and have
universal relevance as they are potentially the building blocks of
larger galaxies.

All the dSph satellites of the Milky Way contain a population of very
old stars and have low mean metallicities similar to those found in
the Galactic stellar halo (Grebel \& Gallagher 2004).  Because the
metallicity of a galaxy increases in time, the most metal-poor stars
must be related to the first stars formed, allowing direct access to
the physical processes and properties of the early Universe. For
example, if dSph were fossils of the pre-reionization era (Gnedin \&
Kravtsov 2006), the first stars in them could have been (partly)
responsible for the reionization of the Universe and the
metal-enrichment of the intergalactic medium (Bromm \& Larson 2004).

{\it Present-day} dSph have been ruled out as the prime contributors
to the Galactic halo on the basis of their stellar populations
(Unavane, Wyse \& Gilmore 1996) and their detailed chemical abundance
patterns (Shetrone, C{\^o}t{\'e} \& Sargent 2001; Venn et al.\ 2004;
Pritzl, Venn \& Irwin 2005).  This is not completely unexpected given
that the surviving dSph have had a Hubble time to evolve as
independent entities under differing conditions. Font et al. (2006)
point out that if the dSph had been accreted at early epochs, they
could still have been the building blocks of hierarchical
models. However, we will show here that even this explanation is not
viable: the high-redshift progenitors of the dSph appear to have been
fundamentally different from the building blocks of the Milky Way.

In this Letter, we study the metallicity distribution of a large
sample of stars in four nearby dSph: Sculptor, Fornax, Sextans and
Carina.  The data presented here were taken using the European
Southern Observatory (ESO) VLT/FLAMES facility in low-resolution (LR)
mode (Pasquini et al. 2002). The DART programme has observed Sculptor,
Fornax and Sextans, while the data for Carina comes from the ESO
archive.

\section{Observations and Analysis}

For each galaxy we derive metallicity estimates as well as accurate
radial velocity measurements for several hundred red giant branch
(RGB) stars located over a large area extending out to the tidal
radius (Tolstoy et al.\ 2004, Koch et al.\ 2006, Battaglia et al.\
2006).  We use RGB stars for our spectroscopic program, since they are
bright, cover a range of ages dating back to the oldest stellar
populations in these galaxies (Stetson, Hesser \& Smecker-Hane 1998),
and (the majority) are believed to contain in their atmospheres an
unpolluted sample of the metallicity of the interstellar medium out of
which they formed.  To identify candidate RGB stars in the dSph we
used ESO Wide Field Imager observations to construct colour-magnitude
diagrams and selected giant branch stars from their well-defined
locii.  We took particular care to acquire a statistically fair sample
by targeting potential RGB members over a wide range in colour and in
spatial location, in order to probe the more evenly distributed
metal-poor component (although we only have a small number of stars
located near the tidal radius). For example, in Sextans we have
obtained spectra for every RGB star candidate over several square
degrees down to $V$=20 ($M_V$=0.6), while for Sculptor more than
$50$\% of the brighter RGB stars out to the notional tidal radius
have been followed up.

We use the CaII triplet region of the RGB star's spectrum to estimate
its metallicity (Cole et al.\ 2004). This indicator requires only low
or intermediate spectral resolution, is based on three absorption
lines around 8500$\AA$, and has been empirically calibrated from
high-resolution abundance studies of stars in globular clusters
(Armandroff \& Da Costa, 1991).

All spectroscopic VLT/FLAMES observations were made in Medusa mode, in
which $\sim$120 fibres are placed over a 25 arcmin diameter field of
view.  The data were all reduced, extracted and wavelength calibrated
using the pipeline provided by the FLAMES consortium (Blecha et
al. 2003).  For sky subtraction, velocity and equivalent width
estimation, we developed our own software which was thoroughly checked
on multiple observations of the same fields taken at different times.
From the equivalent width of the individual CaII triplet lines $EW$,
we estimate the metallicity as (Tolstoy et al.\ 2001):
\begin{equation}
\feh = -2.66 + 0.42 \, [EW_{2+3} + 0.64 \, (V - V_{HB})]
\end{equation}
where $EW_{2+3}$ is the combined equivalent width derived from lines 2
and 3, and $V - V_{HB}$ is the apparent magnitude difference between
the star and the horizontal-branch of the system (i.e.\ a proxy
correction for the effect of surface gravity). We find that for
reliable $\feh$ determinations ($\pm$~0.1 dex) a continuum
signal-to-noise $S/N \geq$ 10 per $\AA$ is required. In what follows,
we consider only those stars satisfying this criterion and also with
estimated radial velocity errors less than 6~km/s.

\begin{figure} 
\begin{center}
\includegraphics[width=83mm]{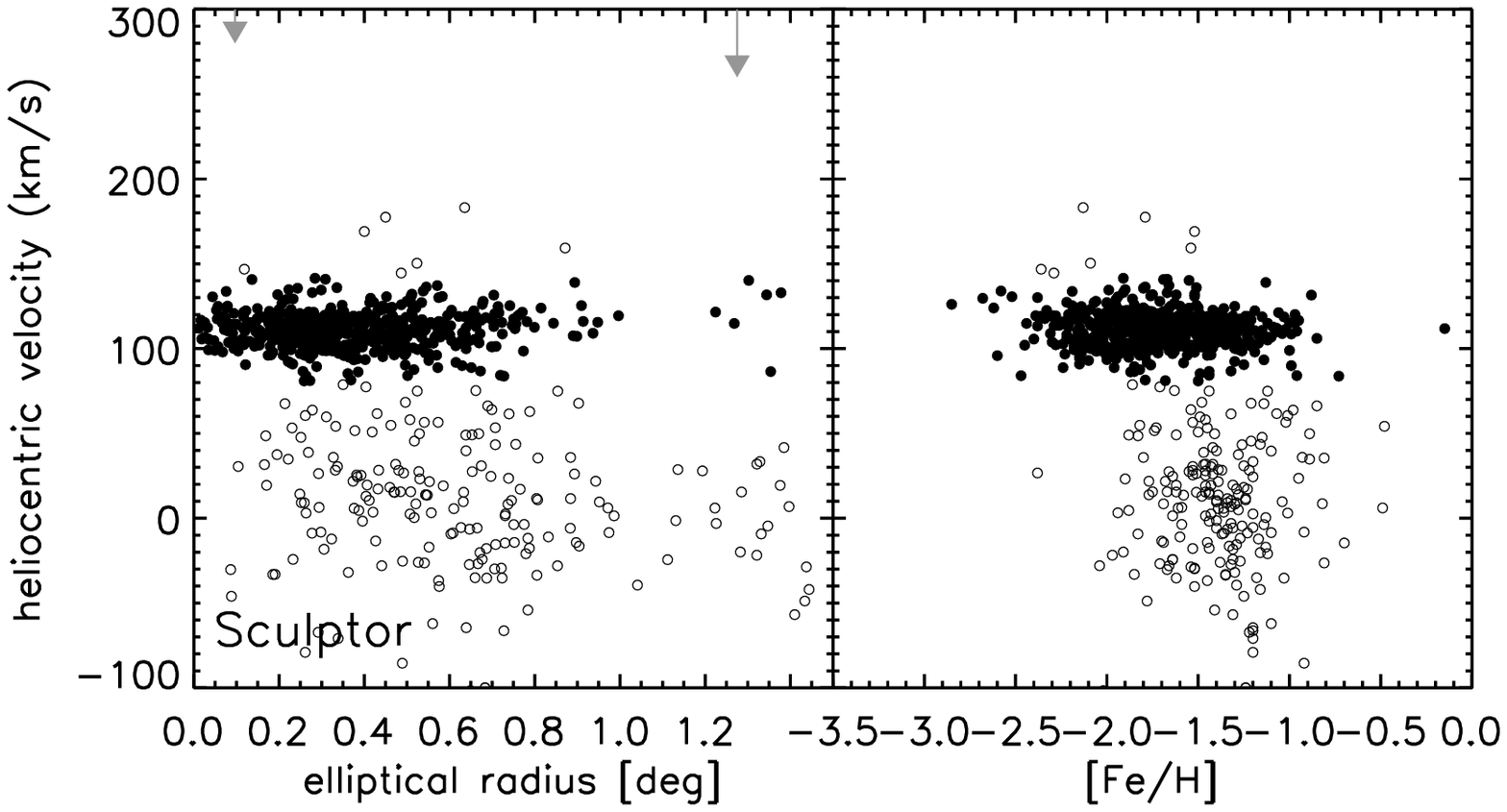}
\includegraphics[width=83mm]{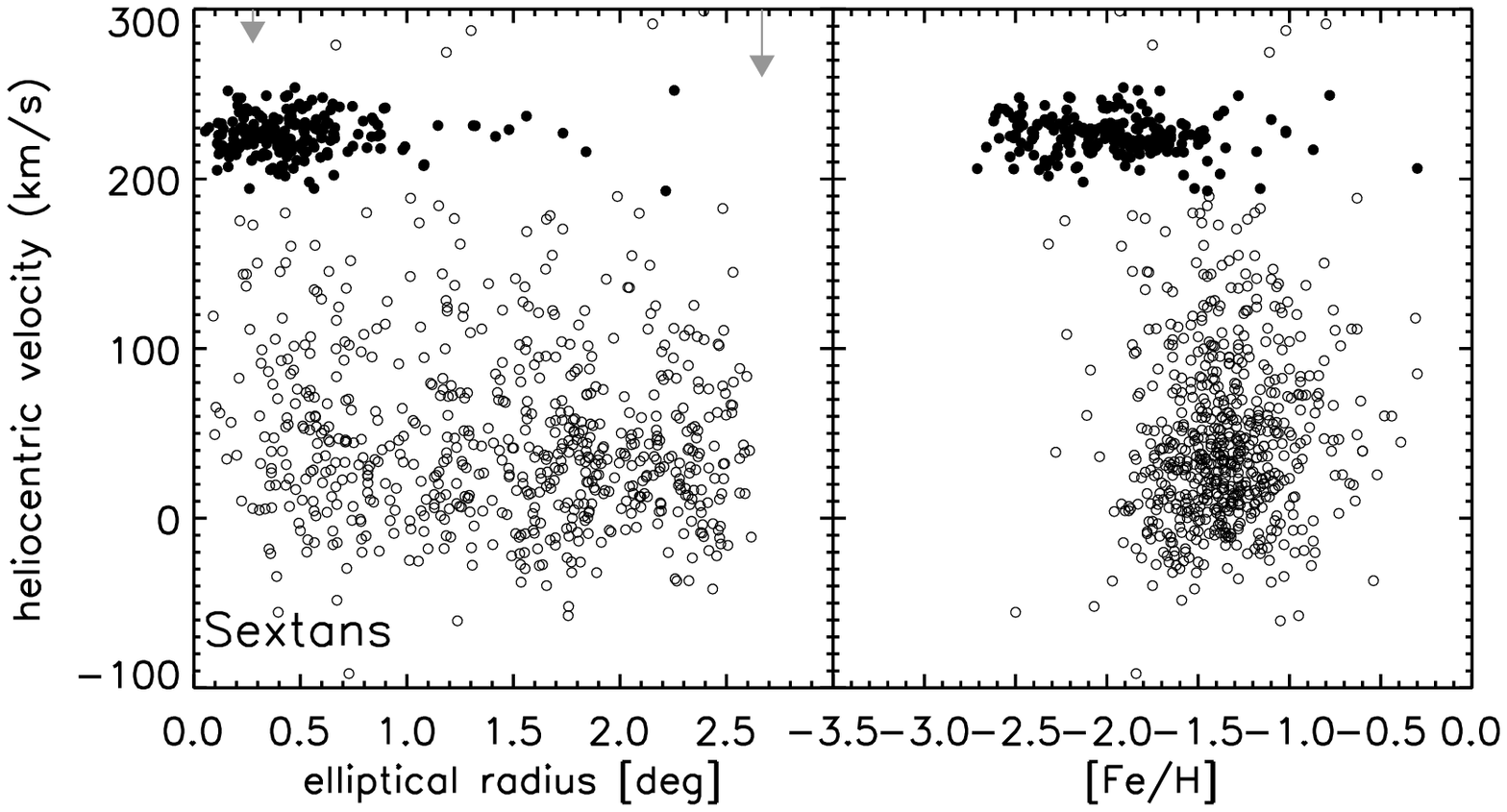}
\includegraphics[width=83mm]{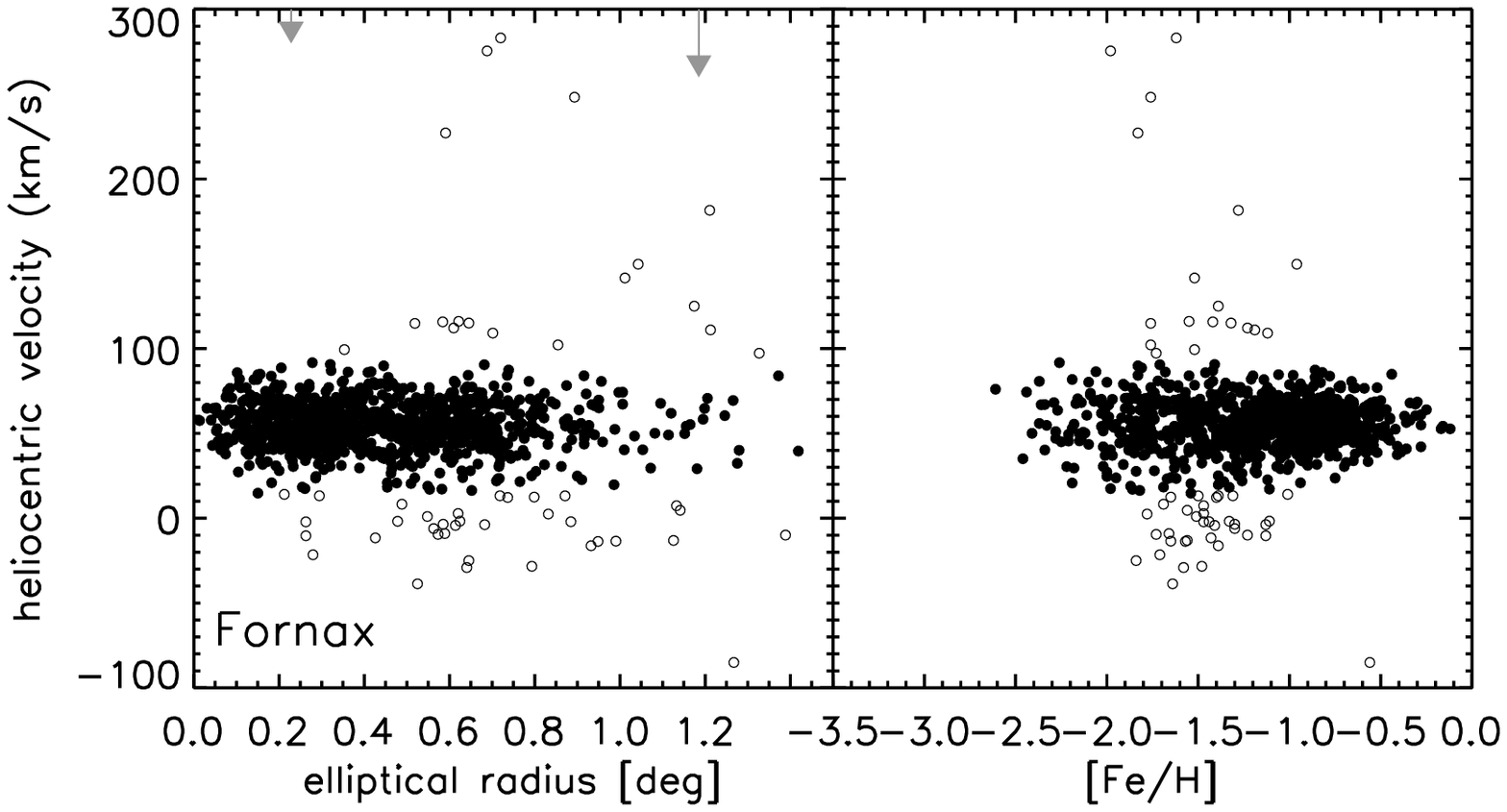}
\includegraphics[width=83mm]{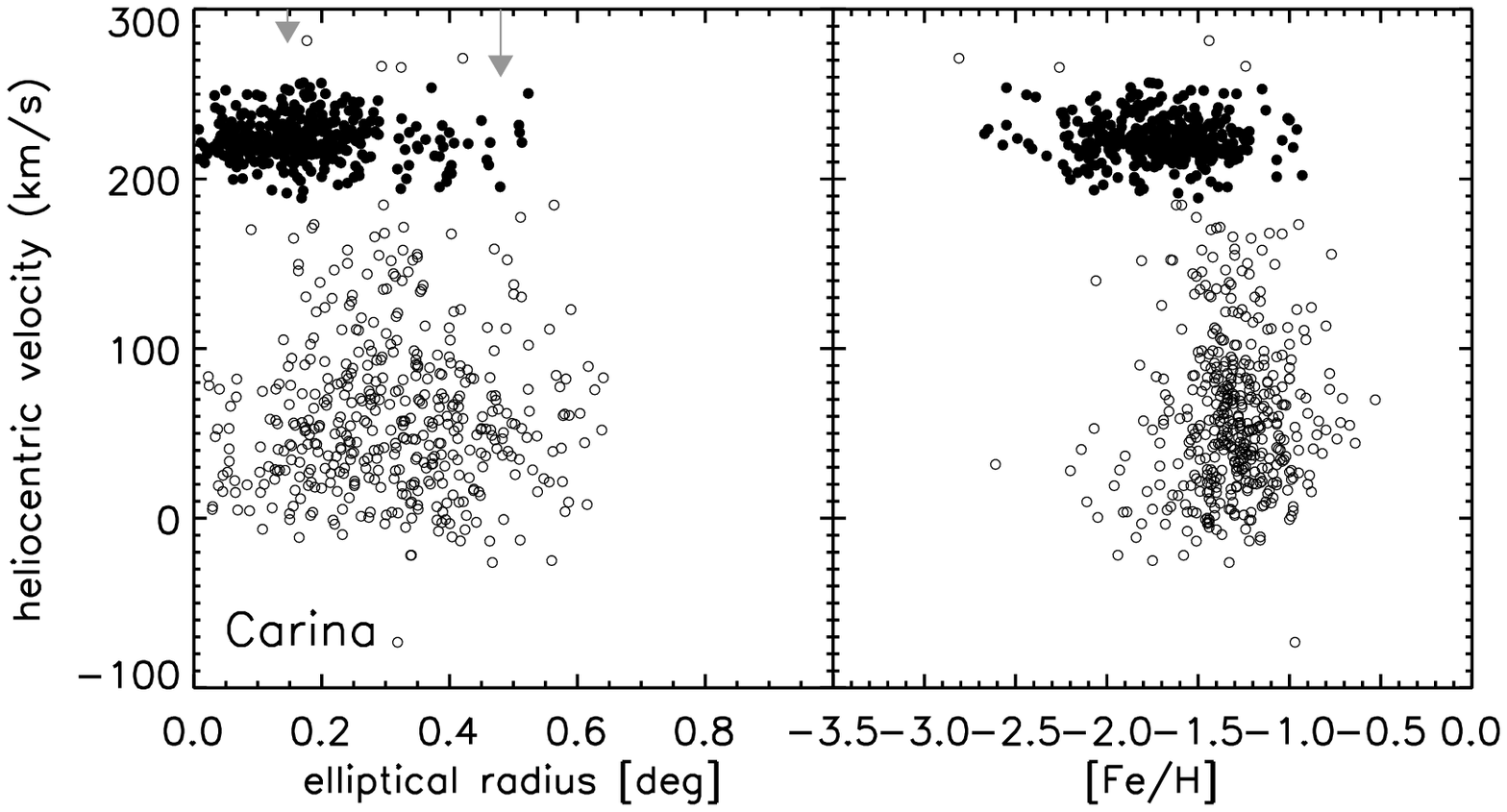}
\end{center}
\caption{Velocities of stars in four dSph (solid symbols), as function
of elliptical radius (left) and metallicity (right). The arrows on the
left panels denote the core and tidal radius of the galaxies. Note
that the foreground dwarf (open symbols) metallicities are token
values; nevertheless the velocity {\it vs.} $\feh$ diagrams are
generally a useful diagnostic of membership.
\label{fig:members}}
\end{figure}

The velocities, metallicities and spatial distribution of the stars in
our survey are shown in Fig.~\ref{fig:members}.  We identify
velocity members of a particular dwarf galaxy by applying a
$\sigma$-clipping procedure, and retaining those stars within
3-$\sigma$ of the converged mean velocity (typically $\pm 35$
km/s). There are 364 member stars in our Carina sample, 933 in Fornax,
513 in Sculptor and 202 in Sextans.

\subsection{The low-metallicity tail}

\begin{figure*}
\includegraphics[height=0.3\textheight]{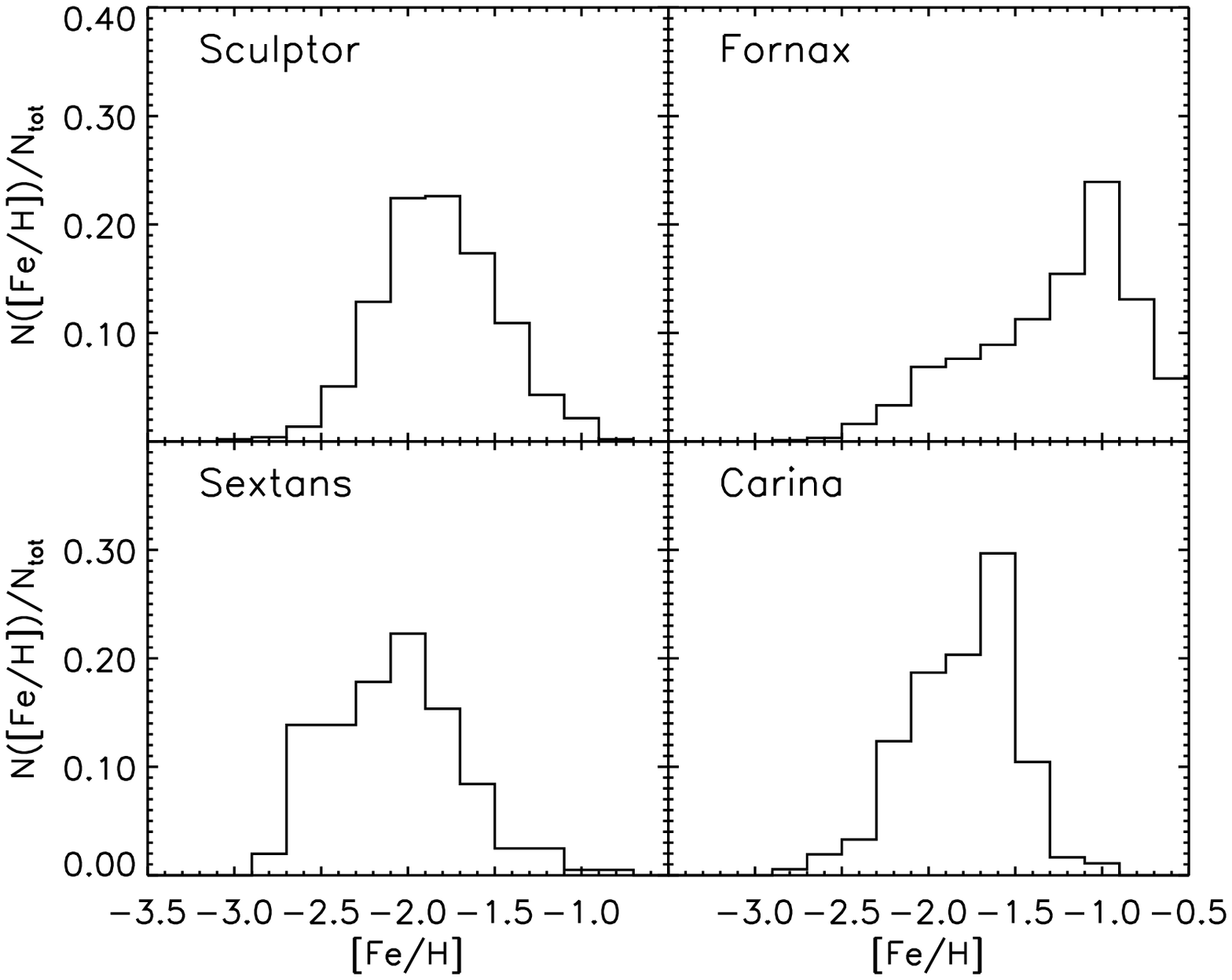}
\includegraphics[height=0.3\textheight]{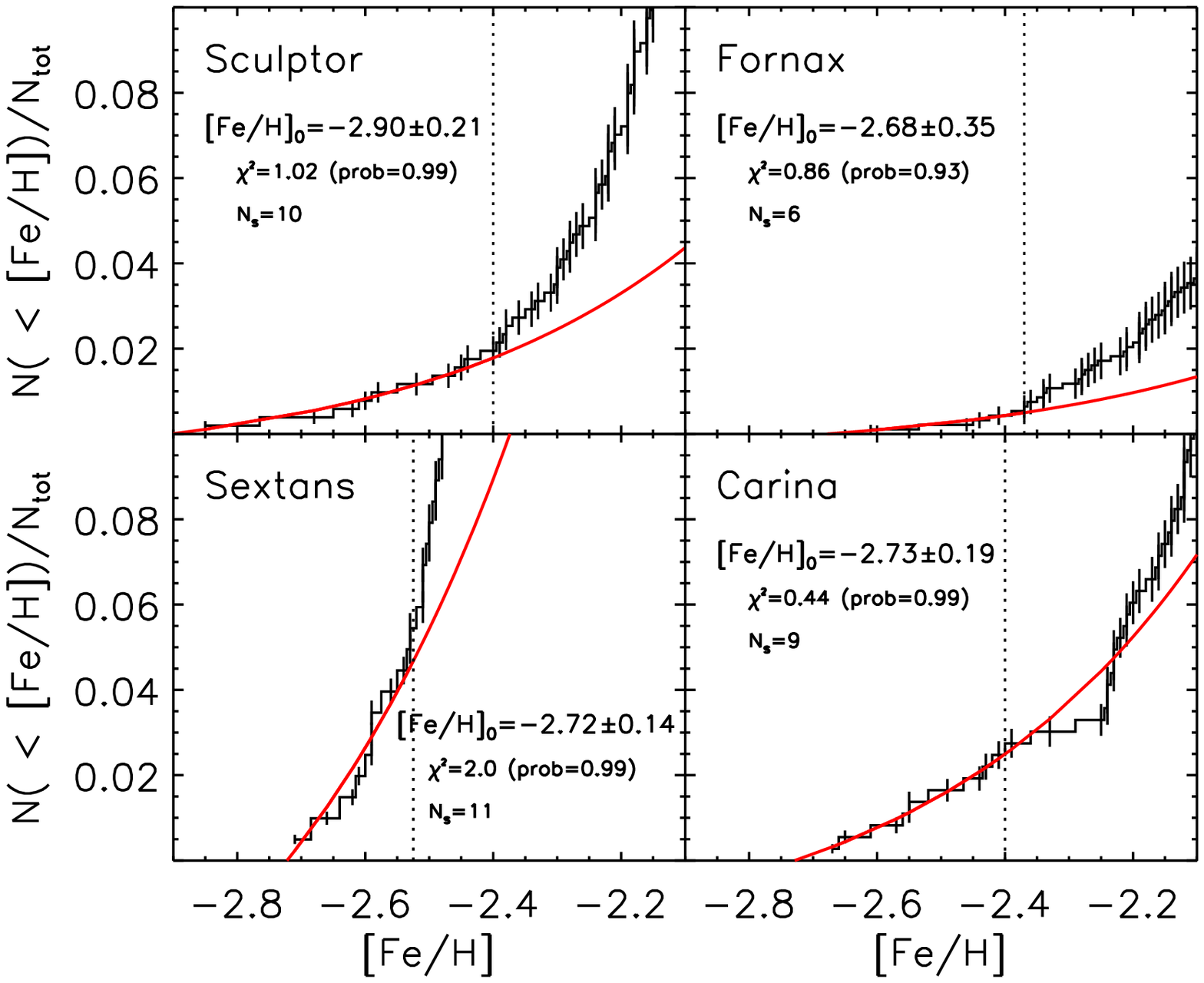}
\caption{{\it Left:} Differential metallicity distribution of the
dSph.  {\it Right:} Cumulative metallicity distributions (error bars
are Poissonian). The solid curve corresponds to the least squares model fit
to the low metallicity tail (defined by the N$_s$ stars located to the
left of the vertical dotted line).
\label{fig:df_dart}}
\end{figure*}

The metallicity distributions of the galaxies are shown in
Fig.~\ref{fig:df_dart}. There is great diversity from system to
system, which reflects their widely different star formation and
chemical enrichment histories (e.g.\ Mateo 1998). There is, however, a
common denominator: contrary to naive expectations, there is a dearth
of stars with $\feh < -3 $ dex.

We have carried out a number of checks to confirm the reality of this
result, and to rule out systematic errors that would prevent us from
finding $\feh< -3$~dex from our LR CaII spectra. Firstly, we have
compared our FLAMES LR CaII measures with FLAMES high-resolution
measures based directly on FeI and FeII lines and find good agreement
(i.e.\ within 0.1 to 0.2 dex), over the range $ -2.5 < \feh < -0.5$
for the Sculptor and Fornax samples (Battaglia et al., in prep).
Secondly, we have followed up with high resolution spectroscopy,
several of the metal-poor objects with $\feh \sim -2.7$ and find
similarly good agreement with our LR metallicity
estimates.  Finally, we have carried out an extensive search for
continuum-only CaII objects in our samples, and found very few
candidates. These have turned out to be nearby subdwarfs or distant
galaxies in follow-up high-resolution spectra with larger wavelength
coverage (Venn et al., in prep).

We now focus on the low-metallicity tail of the distributions
($\feh\lesssim-2.4$~dex). The panels on the right of
Fig.~\ref{fig:df_dart} show the (unbinned) cumulative distribution of
metallicities for the dSph. This distribution is less sensitive to
small numbers statistics than the differential one shown in the
left-handside panels.  The exponential shape of the low-metallicity
tail can be explained in the context of a closed-box model of chemical
evolution with initial enrichment (e.g.\ Pagel 1998). In this model
the number of stars with abundances lower than $Z$ is
\begin{equation}
N(<Z) = A (1 - \exp{\{-(Z - Z_0)/p\}})
\label{eq:1}
\end{equation}
where $A$ depends on the initial gas mass available for star
formation, $Z_0$ is the initial abundance of the gas, and $p$ is the
yield. In the regime considered here, $Z/Z_\odot \sim (10^{-4} -
10^{-2.5})$, and so we can linearize Eq.(\ref{eq:1}) to obtain $N(<Z)
\sim A (Z - Z_0)/p$. If we express $Z = Z_\odot 10^{\feh}$, then
\begin{equation}
N(< \feh )= a \, \exp{(\feh \ln 10)} + b
\end{equation}
where $a = A Z_\odot/p $, and $b = - a \exp{(\feh_0 \ln
10)}$. Therefore the ratio $b/a$ directly depends on the initial
metallicity $\feh_0$ of the gas from which the stars formed.  The grey
curves shown in the panel on the right of Fig.~\ref{fig:df_dart}
correspond to the model fits to the low-metallicity tail of the
distribution. In all cases, we find very good fits to the data, and
are able to obtain a reliable estimate of $\feh_0$.

This analysis shows that the interstellar medium (gas) in each of
these dwarf galaxies had been enriched to $\feh \sim -3$ dex prior to
the earliest star formation episode that led to the present-day
stellar population. It is intriguing that this lowest metallicity is
very similar for all four galaxies, despite their widely different
characteristics. This suggests that the gas had been enriched very
uniformly over a (co-moving) volume of $\sim$ 1 Mpc$^3$ (i.e. that
occupied by the Local Group) very early-on.  This conjecture would
also seem to be supported by the metallicity distribution of Galactic
halo globular clusters which does not extend below $-2.4$ dex
(e.g. Harris 1996).

%\begin{table}[!b]
%\begin{center}
%\begin{tabular}{lcccc}
%\hline
%\hline
%galaxy & $\feh_0$ & $\chi^2$ & $prob$ & $N_{stars}$\\
%\hline
%Sculptor & $-2.90 \pm$ 0.21 & 1.02 & 0.998 & 10 \\
%Fornax & $-2.68 \pm$ 0.35 & 0.86 &  0.931 & 6 \\
%Sextans & $-2.72 \pm$ 0.14 & 2.0  & 0.991 & 11 \\
%Carina & $-2.73 \pm$ 0.19 & 0.44 & 0.999 & 9 \\
%\hline
%\end{tabular}
%\end{center}
%\caption{Least squares fit to the low-metallicity tail of the
%cumulative metallicity distribution of the stars in the dSph.}
%\end{table}

\subsection{Comparison to the Galactic halo}

To establish the relation between the dwarf satellites of our Galaxy
and its putative building blocks, we now compare the metal-poor tail
of the metallicity distributions of the dSph and the Galactic halo.
By focusing on the low-metallicity tail we aim to test whether the
first generations of stars in the dwarfs were analogous to those in
the Galactic building blocks.

The halo metallicity distribution function has been intensively
studied for more than three decades because of its power to constrain
Galaxy evolution models. Ryan \& Norris (1991) were the first to show
that the metal-poor tail extends well below $\feh\sim -3$ dex, on the
basis of a sample of 240 kinematically-selected halo stars, in which
only 5 stars were found to have such low metallicities. Although this
is a small number of objects, this sample size was already smaller
than what is typical for the dSph in the DART program. 

More recently, the HK and the Hamburg/ESO (HES) surveys have yielded a
dramatic increase in the number of metal-poor stars known (Beers \&
Christlieb 2005).  In these surveys, metal-poor candidates are
selected from objective-prism spectra for which the Ca~H and K lines
are weaker than expected at a given $(B-V)$ colour. These stars are
then followed up with medium resolution spectroscopy\footnote{Barklem
et al.\ (2006) have carried out a high-resolution study of a large
subset of stars from the HES survey, and found good agreement with the
Ca~HK metallicity estimates.}. By construction, a bias is introduced at
metallicities $\feh > -2.5$ dex, implying that the shape is only
well-constrained below this value.

The HES survey contains $\sim$40 giants with $\feh\!\!<\!\! -3.0$~dex;
130 stars with $\feh \!\!<\!\! -2.5$~dex and $\sim$400 stars with
$\feh \!<\!  -2.0$~dex (Christlieb, Reimers \& Wisotzki 2004).  We
have 320 stars in the dSph with $\feh \!<\! -2.0$~dex, of which only
29 have $\feh \!<\! -2.5$~dex, and none have $\feh \!<\! -3.0$~dex.
The contrast is stark, but could our inability to find very metal-poor
stars ($\!<\! -3$~dex) in the dSph be an artifact of the sample size?

\begin{figure}
\includegraphics[width=8.5cm]{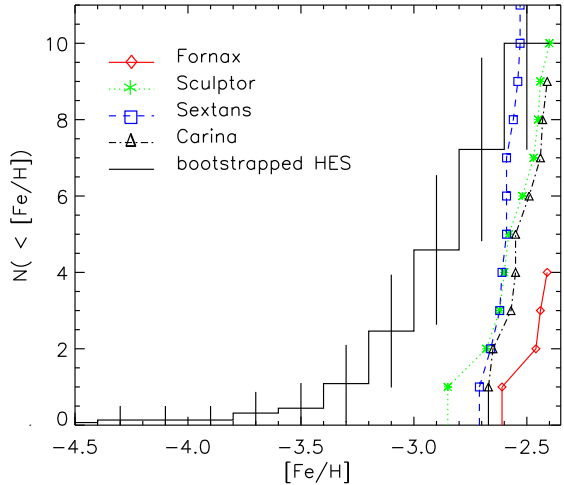}
\caption{Comparison of the cumulative metallicity distributions of the
stars in the mean bootstrapped HES sample and in the dSph.
\label{fig:comp_df}}
\end{figure}

We have quantified the significance of this issue by adopting the
conservative approach of only considering stars with $\feh \!<\!
-2.5$~dex, where the HES survey is most likely to be complete.  We
bootstrapped the HES metallicity distribution below $\feh \sim -2.5$
dex to make random subsets of 10 stars, which is the typical number we
have below $-2.5$ dex for any one of our dSph samples. We then derive
the mean distribution for 1000 such subsets. This is the solid
histogram shown in Fig.~\ref{fig:comp_df}. While it is clear that we
are in the small numbers statistics regime, Fig.~\ref{fig:comp_df}
unambiguously shows that even this bootstrapped distribution is
significantly different from that of the dSph. By means of a
Kolmogorov-Smirnov test, we have quantified the probability that the
bootstrapped HES distribution is consistent with the metallicity
distribution of each individual dwarf galaxy. We find that the
probability is very low in all cases, ranging from $8 \times 10^{-4}$
for Sextans, to $4 \times 10^{-3}$ for Carina and $8 \times 10^{-3}$
for Sculptor and Fornax.

\section{Discussion}

The above analyses show that the tails of the metallicity
distributions of the dSph and the Galactic halo are very different at
highly significant levels. Contrastingly, the metallicity
distributions at low $\feh$ of all dSph are consistent with one
another at the 1-$\sigma$ level; the observed lack of very metal-poor
stars might be considered to be a kind of ``G-dwarf problem" on the
scale of the dwarf spheroidal galaxies.

This implies that any merging, even very early merging, of the
progenitors of the nearby dwarf galaxies as a mechanism for building
up the Galactic halo is ruled out.  The absence of very metal-poor
stars in the dSph shows that the progenitors of the Milky Way and the
dSph must have been different. 

We can think of two possible explanations. In the first scenario the
Galactic building blocks formed from the collapse of high-$\sigma$
density fluctuations in the early Universe (Diemand et al.\ 2005),
while the dwarf satellites would stem from low-$\sigma$ peaks,
predicted to collapse on average at much lower redshifts; e.g., a
1-$\sigma$ density fluctuation of mass $10^8 M_\odot$ collapses at
$z\sim 4$ in a $\Lambda$CDM universe (e.g. Qian \& Wasserburg
2004). It is interesting to note that absorption line spectra towards
quasars show that the intergalactic medium at this redshift has a mean
metallicity of $\feh \sim -3$ dex (Cowie \& Songaila 1998), a value
that is consistent with the lowest metallicity stars found in our
sample. This would mean the oldest stars in the dSph are $\sim 12$ Gyr
old, formed after the Universe was reionized and from a pre-enriched
intergalactic medium.

A second explanation could be that the initial mass function (IMF)
behaved differently in Galactic building blocks and dSph at the
earliest times.  This is a possible interpretation of the
observational fact that the Galactic halo contains some very metal
poor stars with mass $\sim 0.8 M_\odot$ (Christlieb et al. 2002;
Frebel et al. 2005) which are not seen in dSph. For example, in the
case of a bimodal IMF, low-mass stars can form even from zero
metallicity gas (Nakamura \& Umemura 2001). However, this is only
possible if the initial density of the gas is sufficiently large, and
so this would be favoured in high-$\sigma$ peaks collapsing at very
early times.

It may be possible to distinguish between the two scenarios
proposed above through detailed chemical abundance studies.
Unfortunately, the currently available high resolution spectra of
metal-poor stars in dSph is very sparse with only a few published
examples of stars below $-2.5$~dex (e.g. Shetrone et al.\ 2001;
Sadakane et al. 2004; Fulbright, Rich \& Castro 2004) and as such is
still inconclusive.

In both scenarios, the key element is the bias in the galaxy formation
process, while environment would seem to play a less important
role. However, clearly the models need to be explored in much more
detail than the mere outline given here. These efforts should be
supplemented by large observational programs, which should not just
focus directly on the high-redshift distant universe. There is much to
learn about what happened at those early epochs from our own backyard.

\acknowledgements 

This work was partially supported by the Netherlands Organization for
Scientific Research (AH), the Royal Netherlands Academy of Arts and
Sciences (ET), and the U.S. National Science Foundation (AST-0306884,
MDS; AST-0239709, TA). We would like to thank the Aspen Center for
Physics for its hospitality.

\end{document}